\documentclass[a4paper,11pt]{amsart}
\usepackage{amsmath, amsfonts, amssymb, amsthm}
\usepackage{a4wide}
\usepackage{parskip}
\usepackage{enumitem}
\usepackage{xcolor}

\usepackage{hyperref}
\usepackage{booktabs}
\usepackage{caption}
\usepackage{siunitx}
\usepackage{tabularx}
\usepackage{graphicx}
\usepackage{adjustbox}
\usepackage{multirow}
\usepackage{bm}
\usepackage{lscape}
\usepackage{float}
\usepackage{comment}
\usepackage[noadjust]{cite}

\theoremstyle{plain}
\newtheorem{theorem}{Theorem}[section]
\newtheorem{lemma}[theorem]{Lemma}
\newtheorem{proposition}[theorem]{Proposition}
\newtheorem{corollary}[theorem]{Corollary}

\theoremstyle{definition}

\newtheorem{remark}[theorem]{Remark}

\numberwithin{equation}{section}

\usepackage{tikz,xcolor,hyperref}
\newcommand{\Char}{\textnormal{char}}
\newcommand{\Tr}{\textnormal{Tr}_{\mathbb{F}_{q^2}/\mathbb{F}_q}}
\newcommand{\Nr}{\textnormal{N}_{\mathbb{F}_{q^2}/\mathbb{F}_q}}
\newcommand{\Span}{\textnormal{Span}}
\newcommand{\pg}{\textnormal{PG}}
\newcommand{\pgl}{\textnormal{PGL}}
\newcommand{\gl}{\textnormal{GL}}

\begin{document}
\title{Chordal Construction of Column-Twisted Reed--Solomon Codes}
\author{Anuj Kumar Bhagat$^{\ast}$}
\address{Department of Mathematics, Indian Institute of Science Education and Research, Pune, Maharashtra, 411008 India.}
\address{$^{\ast}$Corresponding author}
\author{Krishna Kaipa}
\email{anujkumarbhagat632@gmail.com, kaipa@iiserpune.ac.in}
\begin{abstract}
    A  $[n,k]_q$ Generalized Reed--Solomon (GRS) code  can be represented, up to monomial equivalence, by a projective system consisting of $n$ points on a degree $(k-1)$ rational normal curve (RNC) $C_{k-1}$ in $\pg(k-1,q)$. We use the term \textit{Column-Twisted Reed--Solomon (CTRS)} codes broadly for $[n,k]_q$ codes whose projective system contains at least $(k+2)$  points on a $C_{k-1}$ of $\pg(k-1,q)$. By a chord of $C_{k-1}$, we mean a line which is  either a tangent, a real secant, or an imaginary secant of $C_{k-1}$. Roth and Lempel \cite{RothLempel1989} constructed non-GRS MDS $[n,k]_q$ codes by considering projective systems consisting of $(n-1)$ points on $C_{k-1}$  together with a point on a  tangent line. 
    More recently, Liu et al. \cite{Liu2026} obtained non-GRS $[n, k]_q$ MDS codes by considering projective systems consisting of $(n-2)$ points on $C_{k-1}$  together with a pair of points on a real secant. In the case when $q$ is odd, the length of these codes satisfies  $k+4 \leq n \leq (q+3)/2$. In this article, we construct non-GRS $[n,k]_q$ MDS codes by considering projective systems consisting of $(n-2)$ points on $C_{k-1}$ together with a pair of Galois conjugate points on an imaginary secant. For odd $q$, these codes  have lengths satisfying $k+4 \le n\le (q+5)/2$,  thereby improving both the attainable lengths and dimension ranges of previous TRS and CTRS constructions. In particular, our code  attains the length $(q+5)/2$ for every $3\le k\le (q-3)/2$. We use a geometric argument to establish
    the non-GRS property of CTRS codes for the range $3\le k \le n-4$, while the traditional method of comparison with the   Schur square of GRS codes proves the non-GRS property only for the narrower range $3\le k \le n/2$. We also determine the structure of the dual of any CTRS code.
\end{abstract}
\maketitle
\section{Introduction}\label{Section 1}
    A $[n, k, d]_q$-linear code is a $k$-dimensional subspace of $\mathbb{F}_q^n$ with Hamming distance $d$. The Singleton bound states that every $[n, k, d]_q$-linear code satisfies 
    $d\le n-k+1$. Any linear code that attains the Singleton bound is called a maximum distance separable (MDS) code. MDS codes attain the largest possible minimum distance for a given length and dimension and, consequently, have optimal error-detection and error-correction capabilities. Moreover, the dual of an MDS code is again an MDS code. The fundamental role of MDS codes in coding theory has led to considerable interest in their structural and combinatorial properties. MDS codes are also closely connected to finite geometry, where they correspond to arcs in projective spaces, and have connections with combinatorial designs \cite{Hirschfeld1, macwilliams1977theory}.
    
    For $n\le q+1$, a well-known family of MDS codes of length $n$ over $\mathbb{F}_q$ is given by the Generalized Reed--Solomon (GRS) codes. Most known MDS codes are equivalent to GRS codes, and the construction and classification of MDS codes that are not equivalent to GRS codes remain important and challenging problems in coding theory. Such codes are referred to as non-GRS MDS codes. It is well known that every $[n,k]$-MDS code with $k\le 2$ is equivalent to a GRS code. Since the dual of a GRS code is again a GRS code, non-GRS MDS codes can occur only when $3\le k\le n-3.$

    One of the first constructions of  non-GRS MDS codes was given by Glynn in \cite{glynn1986non}, who constructed an $[10, 5]$ non-GRS MDS code over $\mathbb{F}_9$. Subsequently, Roth--Lempel \cite{RothLempel1989} constructed non-GRS MDS codes over $\mathbb{F}_q$ by extending Reed--Solomon (RS) codes by one coordinate. These codes are now referred to as Roth--Lempel codes. For odd $q$, there exist $[k+3, k]$ MDS Roth--Lempel codes for $4\le k\le (q-1)/2$. A complete classification of non-GRS MDS codes of dimension $3$ that are MDS extensions of GRS codes was given in \cite{kaipa2017deep}. Beelen et al. introduced twisted GRS codes in \cite{beelen2017twisted, beelen2022twisted} by modifying the rows of a generator matrix of GRS codes. For odd $q$, they constructed non-GRS MDS codes of length $(q+1)/2$ over $\mathbb{F}_q$. More recently, Liu et al. \cite{Liu2026} introduced Column-Twisted Reed--Solomon (CTRS) codes by extending RS codes by one or two coordinates. For odd $q$, they constructed non-GRS MDS codes of length $(q+3)/2$ over $\mathbb{F}_q$, thereby increasing the achievable length compared with the previously known twisted GRS constructions. Various other constructions of non-GRS MDS codes have been studied; see, for example, \cite{Oggier2026, abdukhalikov2026some, chen2023many, li2025cyclic, jin2025new} and the references therein. Variants of Roth--Lempel codes were considered in \cite{li2025new, sui2027new}.

    In this article, we study extensions of GRS codes by one or two coordinates whose projective system consists of $n$ points on an RNC and at most two points on a chord of the RNC (for the definition of a chord, see Section \ref{Section 2}).
    For example, the projective systems of $[n,k]_q$ Roth--Lempel codes consist of $(n-1)$ points on $C_{k-1}$ and one  point on a tangent line to $C_{k-1}$. Similarly, the projective systems of the $[n,k]_q$ CTRS  codes obtained in \cite{Liu2026} consist of $(n-2)$ points on $C_{k-1}$  together with a pair of points on a  real secant. In this work, we introduce $[n,k]_q$ codes whose projective systems consist of $(n-2)$ points on $C_{k-1}$  together with a pair of Galois conjugate points on an imaginary secant of $C_{k-1}$. We derive necessary and sufficient conditions for such codes to be MDS and provide explicit constructions. We use the Schur-square technique to establish the non-GRS property of these codes. Specifically, by comparing the dimension of the Schur square of an MDS CTRS code with that of a GRS code, we show that these codes are non-GRS whenever these dimensions differ. However, this approach establishes the non-GRS property of $[n, k]_q$ CTRS codes only for $3\le k \le \frac{n}{2}.$ To overcome this limitation, we use a geometric argument to show that any $[n,k]_q$ MDS CTRS code is non-GRS for $3\le k \le n-4$. In particular, for odd $q$, we construct non-GRS MDS codes of length $(q+5)/2$ and dimension $k$ for every $3 \le k \le (q-3)/2$. We also determine the structure of the dual of any code in this class and show that its dual is a (row)-twisted GRS code.

    The rest of the article is organized as follows. In Section \ref{Section 2}, we provide the necessary preliminaries. In Section \ref{Section 3}, we introduce and study Column-Twisted Reed--Solomon codes. In Section \ref{section 4}, we investigate the duals of CTRS codes.
    
\section{Preliminaries}\label{Section 2}
Throughout this article, $\mathbb{F}_q$ denotes the finite field of order $q$. For the quadratic extension $\mathbb{F}_{q^2}/\mathbb{F}_q,$ the \textit{trace} and the \textit{norm} maps are defined, respectively, by
$$
\begin{aligned}
    \Tr:\mathbb{F}_{q^2}
    &\longrightarrow \mathbb{F}_q,\\
    \alpha
    &\longmapsto\alpha+\alpha^q,\ \forall\ \alpha\in \mathbb{F}_{q^2},
\end{aligned}
$$
and
\[
\begin{aligned}
    \Nr:\mathbb{F}_{q^2}^{\times}
    &\longrightarrow \mathbb{F}_q^{\times},\\
    \alpha
    &\longmapsto\alpha^{q+1},\ \forall\ \alpha\in \mathbb{F}_{q^2}^{\times}.
\end{aligned}
\]
The norm map is a surjective group homomorphism from $\mathbb{F}_{q^2}^{\times}$ to $\mathbb{F}_q^{\times}$ with kernel 
$$
\ker (\Nr) :=\{\alpha\in \mathbb{F}_{q^2}^{\times} : \alpha^{q+1}=1\}.
$$

The projective space $\pg(k-1,q)$ is the set of all one-dimensional subspaces of $\mathbb{F}_q^k$, or in other words, equivalence classes of $(\mathbb{F}_q^k\setminus \{\bm{0}\})/\sim$,  
where $\sim$ is the equivalence relation defined by $\bm{u}\sim\bm{v}\iff \bm{u}=\lambda \bm{v}$ for some $\lambda\in \mathbb{F}_q^\times.$ We continue to denote the equivalence class of a non-zero vector $\bm{u}=(u_0,u_1,\dots,u_{k-1})$ by the same symbol $(u_0,u_1,\dots,u_{k-1})$ when the appropriate space  is clear from context.

For $t\in \mathbb{F}_q\cup\{\infty\},$ we define 
    $$
        c_{k-1}(t):= \begin{cases}
            (1,t,t^2,\dots,t^{k-1})^\top, & t\ne \infty\\
            (0,\dots,0, 1)^\top, & t=\infty.
        \end{cases}
    $$ 
    and    
    $$
        c_{k-1}'(t):=  \begin{cases}
            (0, 1, 2t, \dots,(k-1)t^{k-2})^\top, & t\ne \infty\\
            (0, 0, \dots, 1, 0)^\top, & t=\infty.
        \end{cases}
    $$

    Let 
    \begin{equation}
        \mathbb{F}_q[x]_{<k}:=\{f(x)\in\mathbb{F}_q[x]: \deg f(x)< k\}\cup \{0\}.
    \end{equation}
    For $k\le n\le q+1,$ let $\bm{x}:=(x_1, x_2, \dots, x_n)$ be an ordered $n$-tuple of distinct elements of $\mathbb{F}_q\cup \{\infty\}$ and $\bm{v}=(v_1, \dots, v_n)\in (\mathbb{F}_q^{\times})^n$. A \textit{Generalized Reed--Solomon (GRS) code} of length $n$ and dimension $k$ with evaluation set $\bm{x}$ is defined by
    \begin{equation} \label{eq:def_GRS}
        \text{GRS}_k(\bm{x}, \bm{v}):=\left\{(v_1f(x_1), v_2f(x_2), \dots, v_nf(x_n)): f(x)\in\mathbb{F}_q[x]_{<k}\right\},
    \end{equation}
    where $f(\infty)$ is defined as the coefficient of $x^{k-1}$ in $f(x).$
    
    It is well known that every GRS code is an MDS code. Moreover, the dual of $\text{GRS}_k(\bm{x}, \bm{v})$ is $\text{GRS}_{n-k}(\bm{x}, \bm{w})$, where 
    \begin{equation}\label{eqn: dual of GRS}
        w_i=
        \begin{cases}
        v_i^{-1}
        \!\!\!\!\!\!\displaystyle\prod_{\substack{j=1\\ j\ne i,\ x_j\ne\infty}}^n
        \!\!\!\!\!\!(x_i-x_j)^{-1}, & \text{if } x_i\ne\infty,\\[3mm]
        -v_i^{-1}, & \text{if } x_i=\infty.        \end{cases}
    \end{equation}
    Up to monomial equivalence, a GRS code with evaluation set $\bm{x}$ can be represented by the generator matrix
    \begin{equation*}
            \begin{bmatrix}
                c_{k-1}(x_1)  & c_{k-1}(x_2) & \cdots & c_{k-1}(x_n)
            \end{bmatrix}.
    \end{equation*}
   
    The image of the Veronese embedding
    \begin{align*}
        \nu_{k-1}: \pg(1, q)&\hookrightarrow \pg(k-1, q)\\
        \nu_{k-1}(t)&=c_{k-1}(t).
    \end{align*}
    is  the \textit{standard rational normal curve} of degree $(k-1)$, denoted by $C_{k-1}$. In general, a \textit{rational normal curve (RNC)} of degree $(k-1)$ in $\pg(k-1, q)$ is $g(C_{k-1})$ for some  $g\in \pgl_k(\mathbb{F}_q)$. By a point of $C_{k-1}$, over an extension field $\mathbb{F}_{q^m}$, we mean $\nu_{k-1}(t)$ for $t \in \pg(k-1,q^m)$.
    Clearly, $C_{k-1}$ contains exactly $(q^m+1)$ points of $\pg(k-1,q^m)$. A \textit{tangent line} to $C_{k-1}$ at $c_{k-1}(a)$ is the projective line joining $c_{k-1}(a)$ and $c_{k-1}'(a)$. A \textit{real secant} of $C_{k-1}$ is the line joining $c_{k-1}(a)$ and $c_{k-1}(b)$, where $a, b\in\mathbb{F}_q \cup \{\infty\}$ are distinct. An \textit{imaginary secant} of $C_{k-1}$ is the line joining $c_{k-1}(\alpha)$ and $c_{k-1}(\alpha^q)$, where $\alpha\in \mathbb{F}_{q^2}\setminus \mathbb{F}_q$. Following \cite[p.231]{Hirschfeld3}, by a \textit{chord} \label{defn: chord} of $C_{k-1}$, we mean either a tangent line, a real secant  or an imaginary secant of $C_{k-1}$. 
    
    The \textit{projective system} associated to a non-degenerate $[n,k]_q$-code $\mathcal C$, is  the  multiset $\mathcal P$ of $\pg(k-1,q)$ of size $n$ consisting of points represented by the columns of a generator matrix of $\mathcal C$. 
    We say two projective systems $\mathcal P$ and $\mathcal P'$ are\textit{ projectively equivalent} if there is a $g \in \pgl_{k}(\mathbb F_q)$ such that $\mathcal P'=g(\mathcal P)$.
    Clearly, there is a bijective correspondence  between monomial equivalence classes of non-degenerate $[n,k]_q$ codes on one hand and  projective systems of size $n$ in $\pg(k-1,q)$ up to projective equivalence on the other hand. In particular, the projective system associated to a monomial equivalence class of a $[n,k]_q$ GRS code, can be represented by $\mathcal{P}=\{\nu_{k-1}(x_1), \dots, \nu_{k-1}(x_n)\}$. 
    

    The projective general linear group $\pgl_2(\mathbb{F}_q)$ acts naturally on the projective line $\pg(1,q)=\mathbb{F}_q\cup\{\infty\}$ by fractional linear transformations. More precisely, an element $g\in\pgl_2(\mathbb{F}_q)$ represented by a matrix $(\begin{smallmatrix} 
    a & b\\ c & d 
    \end{smallmatrix})\in\gl_2(\mathbb{F}_q)$ acts as
    $$
        t\mapsto g(t)=\frac{c+dt}{a+bt},
    $$
    with the convention that $g(t)=\infty$ when $a+bt=0$ and $g(\infty)=\infty$, if $b=0$, while $g(\infty)=d/b$ if $b\ne 0$. This action of $\pgl_2(\mathbb{F}_q)$ on $\pg(1,q)$ induces an action on the standard RNC $C_{k-1}$. More precisely, there exists an injective group homomorphism $g\in\pgl_2(\mathbb{F}_q)\mapsto g^{[k]}\in\pgl_k(\mathbb{F}_q)$ such that $g^{[k]}\big(\nu_{k-1}(t)\big)=\nu_{k-1}\big(g(t)\big)$, for every $t\in\pg(1,q)$(see, for example, \cite[Propositions 2.5, 2.6]{Beelen_Counting_grs}). Thus, $g^{[k]}$ maps $C_{k-1}$ onto itself, and GRS codes with evaluation sets $\bm{x}$ and $g(\bm{x})$ are monomially equivalent for every $g\in\pg(1,q)$.
    
    Throughout this article, we always assume that $k<q$ so that the RNC $C_{k-1}$ has at least $k+2$ points.

    A collection $\mathcal{P}$ of $n$ points in $\pg(k-1,q)$ is said to be an \textit{$n$-arc} if no $k$ points of $\mathcal{P}$ lie on a hyperplane of $\pg(k-1, q)$. Clearly, projective equivalence classes of $n$-arcs of $\pg(k-1,q)$ correspond to monomial equivalence classes of $[n,k]_q$ MDS codes, and in particular, projective equivalence classes of $n$-arcs contained in a degree $(k-1)$ RNC of $\pg(k-1,q)$ correspond to  monomial equivalence classes of $[n,k]_q$ GRS codes. Any MDS code that is not monomially equivalent to a GRS code is referred to as a \textit{non-GRS MDS code}. For $k\in \{1, 2\},$ any $[n,k]_q$-MDS code is equivalent to a GRS code. Since the dual of a GRS code is again a GRS code, a non-GRS $[n,k]_q$-MDS code can only occur for $3\le k \le n-3.$
    \begin{lemma}\cite{Beelen_Counting_grs}\label{thm: Uniqueness of RNC}
        Let $k<q$. A $(k+2)$-arc of $\pg(k-1, q)$ is contained in a unique degree-$(k-1)$ RNC.
    \end{lemma}
    \begin{corollary}\label{corr: geometric proof idea}
        Let $k+2\le n\le q+1$, and let $\bm{x}:=(x_1, x_2, \dots, x_n)$ be an ordered $n$-tuple of distinct elements of $\mathbb{F}_q\cup \{\infty\}.$ Let $\bm{u}\in \mathbb{F}_q^k.$ Then the code generated by $$G(\bm{x}, \bm{u})=\begin{bmatrix}
            c_{k-1}(x_1) & c_{k-1}(x_2) & \cdots & c_{k-1}(x_n) & \bm{u}
        \end{bmatrix}$$ is a $[n+1, k]$ non-GRS code if and only if  $\bm{u}$ does not represent a point of $C_{k-1}.$
    \end{corollary}
    \begin{proof}
     If $\bm{u}$ represents a point of $C_{k-1}$, then clearly the code generated by $G(\bm{x}, \bm{u})$ is GRS. Conversely, if the code generated by $G(\bm{x}, \bm{u})$ is GRS, then the projective system represented by the columns of $G(\bm{x}, \bm{u})$ is contained in a rational normal curve, say $C_{k-1}'$. In particular, both $C_{k-1}$ and $C'_{k-1}$ are degree-$(k-1)$ RNC's containing the $n$-arc $\mathcal P =  \{c_{k-1}(x_1), \dots, c_{k-1}(x_n)\}$. Since $n \geq k+2$, we see that $C_{k-1}=C'_{k-1}$ by Lemma \ref{thm: Uniqueness of RNC}. Consequently, $\bm{u}$ represents a point of $C_{k-1}$.
    \end{proof}
    For $k\le n\le q,$ let $\bm{x}:=(x_1, x_2, \dots, x_n)$ be an ordered $n$-tuple of distinct elements of $\mathbb{F}_q$. For $\lambda\in \mathbb{F}_q,$ a \textit{Roth--Lempel code} $\text{RL}_k(\bm{x}, \lambda)$ \cite{RothLempel1989} of dimension $k$ over $\mathbb{F}_q$ with evaluation set $\bm{x}$ is the code generated by the matrix
    \begin{equation}\label{eqn: Roth-Lempel defn}
        \begin{bmatrix}
            c_{k-1}(x_1) & \cdots & c_{k-1}(x_n) & c_{k-1}(\infty) &\lambda c_{k-1}(\infty)+c_{k-1}'(\infty)
        \end{bmatrix}.
    \end{equation}
    The projective system associated with the monomial equivalence class of a Roth--Lempel code consists of $(n+2)$ distinct points $\{P_1, \dots, P_{n+2}\}$ of $\pg(k-1,q)$, where $P_1, \dots, P_{n+1}$ are on the  RNC $C_{k-1}$ and $P_{n+2}$ lies on the  tangent line to $C_{k-1}$ at a point $P \in \{P_1, \dots P_{n+1}\}$. The code $\text{RL}_k(\bm{x}, \lambda)$ is MDS if and only if $\lambda\ne \sum_{i\in I} x_i$, for every subset $I\subset [n]$ of size $k-1$. For odd $q$, Roth and Lempel in \cite{RothLempel1989} constructed non-GRS $[k+3, k]$ MDS codes, for $4\le k\le \frac{q-1}{2}.$
    
    The Schur product of $\bm{x}=(x_1, x_2, \dots, x_n)\in\mathbb{F}_q^n$ and $\bm{y}=(y_1, y_2, \dots, y_n)\in\mathbb{F}_q^n$ is defined by $\bm{x}\star \bm{y}:=(x_1y_1, x_2y_2, \dots, x_ny_n).$ For two linear codes $\mathcal{C}$ and $\mathcal{D}$ over $\mathbb{F}_q$ of the same length, their \textit{Schur product} is defined by
    \begin{equation}
        \mathcal{C}\star\mathcal{D}:=\Span_{\mathbb{F}_q}\{\bm{c}\star\bm{d}: \bm{c}\in\mathcal{C},\bm{d}\in\mathcal{D}\}.
    \end{equation}
    In particular, $\mathcal{C}^{\star 2}:=\mathcal{C}\star \mathcal{C}$ is called the \textit{Schur square} of $\mathcal{C}.$ If $\mathcal{C}$ and $\mathcal{D}$ are monomially equivalent, then so are their Schur squares $\mathcal C^{\star 2}$ and $\mathcal D^{\star 2}$. The Schur square of the code generated by $[c_{k-1}(x_1) \ c_{k-1}(x_2) \ \cdots \ c_{k-1}(x_n)]$ is the rowspace of $[c_{2k-1}(x_1) \ c_{2k-1}(x_2)\ \cdots \ c_{2k-1}(x_n)]$, which is clearly GRS if $2k-1<n$ and it equals $\mathbb{F}_q^n$ if $2k-1\ge n$, which is $[n,n]_q$-GRS code. Therefore, the Schur square of a $[n,k]_q$-GRS code is a $[n, \min\{2k-1, n\}]_q$-GRS code. This leads us to the next remark.
    \begin{remark}\label{remark: non-RS condition}
     If $\mathcal{C}$ is a $[n,k]_q$-MDS code with $3\le k\le {n}/{2}$ such that $\dim \mathcal{C}^{\star 2}\ge 2k,$ then $\mathcal{C}$ is a non-GRS MDS code.
\end{remark}
For $\{x_1, x_2, \dots, x_n\}\subseteq \mathbb{F}_q, $ the matrix
    \begin{equation}
        V(x_1, x_2 \dots, x_n):=\begin{bmatrix}
            c_{n-1}(x_1) & c_{n-1}(x_2) & \cdots & c_{n-1}(x_n)
        \end{bmatrix}
    \end{equation}
    is called a \textit{Vandermonde matrix} associated with $x_1, x_2, \dots, x_n.$
The determinant of the Vandermonde matrix $V(x_1, x_2 \dots, x_n)$ is denoted by $\Delta_n(x_1, x_2, \dots, x_n).$ It is well-known that
\[
    \Delta_n(x_1, x_2, \dots, x_n):=\underset{1\le i<j\le n}{\prod}(x_j-x_i).
\]
We note that Vandermonde matrix $V(x_1, \dots, x_{n-1}, \infty)$ has determinant $\Delta_{n-1}(x_1, \dots, x_{n-1})$.\\

Let $\{x_1, x_2,\dots,x_n\}\subseteq\mathbb F_q$. The \textit{elementary symmetric polynomials} $e_i(x_1,\dots,x_n)$ are defined by the following identity in the polynomial ring $\mathbb{F}_q[X]$:
\begin{equation}\label{eqn: defn elementary symm}
    \prod_{j=1}^n(X-x_j)
    =\sum_{i=0}^n (-1)^i e_i(x_1,\dots,x_n) X^{n-i}
\end{equation}
where $e_0(x_1,\dots,x_n)=1$ and $e_i(x_1,\dots,x_n)=0$ for $i>n$.

\section{Column-Twisted Reed--Solomon Codes}\label{Section 3}
\subsection{CTRS Codes via Points on Tangent Lines}
    In this subsection, we consider $[n+1,k]_q$ and $[n+2,k]_q$ CTRS codes whose projective systems consists of $n$ points on $C_{k-1}$ and the remaining point(s) on a tangent line to $C_{k-1}$.
    
    For $1\le n\le q+1,$ let $\bm{x}:=(x_1,x_2, \dots, x_{n})$, be an ordered $n$-tuple of distinct elements of $\mathbb{F}_q\cup \{\infty\}$, and let $\lambda\in \mathbb{F}_q$. For $a\in (\mathbb{F}_q\cup\{\infty\})\setminus \{x_i: 1\le i\le n\}$, define
    \begin{equation}
        G_k^{(1)}(\bm{x}, a, \lambda):=\begin{bmatrix}
            c_{k-1}(x_1) & c_{k-1}(x_2) & \cdots & c_{k-1}(x_n) & \lambda c_{k-1}(a)+ c_{k-1}'(a)
        \end{bmatrix},\ \text{and}
    \end{equation}
    \begin{equation}
        G_k^{(2)}(\bm{x}, a, \lambda):=\begin{bmatrix}
            c_{k-1}(x_1) & c_{k-1}(x_2) & \cdots & c_{k-1}(x_n) & c_{k-1}(a) & \lambda c_{k-1}(a)+ c_{k-1}'(a)
        \end{bmatrix}.
    \end{equation}
    For $\lambda_1, \lambda_2 \in \mathbb{F}_{q}$ with $\lambda_1\ne \lambda_2$, and $a\in (\mathbb{F}_q\cup\{\infty\})\setminus \{x_i: 1\le i\le n\}$, define
    \begin{equation} \label{eq:2point_roth_lempel}
        G_k(\bm{x}, a, \lambda_1, \lambda_2):=\begin{bmatrix}
            c_{k-1}(x_1) & \cdots & c_{k-1}(x_n) & \lambda_1c_{k-1}(a)+ c_{k-1}'(a) & \lambda_2 c_{k-1}(a)+ c_{k-1}'(a)
        \end{bmatrix}.
    \end{equation}
Applying the projective transformation $g(t)=1/(t-a)$ maps the tangent line to the RNC at $c_{k-1}(a)$ to the tangent line at $c_{k-1}(\infty)$. Therefore, up to monomial equivalence, we may assume $a=\infty$. Thus, the code generated by $G_k^{(2)}(\bm{x},a,\lambda)$ is monomially equivalent to a Roth--Lempel code \eqref{eqn: Roth-Lempel defn}, and the code generated by $G_k^{(1)}(\bm{x}, a, \lambda)$ is monomially equivalent to a punctured Roth--Lempel code.


The projective system of the code generated by $G_k(\bm{x},a,\lambda_1,\lambda_2)$ consist of $n$ points on $C_{k-1}$ together with two distinct points on the tangent line at $c_{k-1}(a)$. As above, we may assume $a=\infty$ up to monomial equivalence. 
The subgroup of $\pgl_2(\mathbb{F}_q)$ fixing $\infty$ is  
 \[ \mathcal{G}_\infty:=\{g(t)=c+dt: c\in \mathbb{F}_q, d\in \mathbb{F}_q^{\times}\}.\] 
 The affine tangent line at $c_{k-1}(\infty)$ is 
 $\{ s c_{k-1}(\infty)+c'_{k-1}(\infty) \colon s \in \mathbb F_q\}$.
 The action of an element $g(t)=(c+dt)$  of $\mathcal G_\infty$  on the  affine tangent line  is given by 
 $g(t) \cdot s = (k-1)c+ds$, in terms of the parameter $s$ on  the line. If $k\equiv 1\pmod p$ where $p=\Char(\mathbb{F}_q)$, then the action reduces to $g(t) \cdot s = d s$. Consequently, there are exactly two $\mathcal{G}_{\infty}$-orbits on the affine tangent line, namely
$$ \{c'_{k-1}(\infty)\}
\qquad\text{and} \qquad
\{ c'_{k-1}(\infty) + s\, c_{k-1}(\infty) \colon  s \in \mathbb{F}_q^{\times} \}.
$$
Therefore, up to monomial equivalence, the generator matrices in \eqref{eq:2point_roth_lempel} can be taken to be  either
\[ 
G_1=
\begin{bmatrix}
    c_{k-1}(y_1) & \cdots & c_{k-1}(y_n) & c_{k-1}'(\infty) & c_{k-1}(\infty)+c_{k-1}'(\infty)
\end{bmatrix},
 \]
or 
\[ 
G_2=
\begin{bmatrix}
    c_{k-1}(y_1) & \cdots & c_{k-1}(y_n) & c_{k-1}(\infty)+c_{k-1}'(\infty) & \lambda c_{k-1}(\infty)+c_{k-1}'(\infty)
\end{bmatrix},
\]
according as $\lambda_1 \lambda_2$ is zero or non-zero. Here $y_i$ are distinct elements of $\mathbb{F}_q$ and $\lambda\in\mathbb{F}_q^{\times}\setminus\{1\}$.


If $k\not\equiv1\pmod p,$ then the induced action of $g(t)=(c+dt)$ on the parameter $s$ of the affine tangent line is $g(t) \cdot s = (k-1)c+ds$. Since $k-1\neq0$ in $\mathbb{F}_q$, we see that $\mathcal G_\infty$ acts doubly-transitively on this line.
Consequently, since $\lambda_1\neq\lambda_2$, we may assume that $(\lambda_1,\lambda_2)=(0,1)$.
 Therefore,  then  up to monomial equivalence, the generator matrices in \eqref{eq:2point_roth_lempel} can be taken to be 
\[
G_3=
\begin{bmatrix}
c_{k-1}(y_1)&\cdots&c_{k-1}(y_n)&
c_{k-1}'(\infty)&
c_{k-1}(\infty)+c_{k-1}'(\infty)
\end{bmatrix},
\]
where $y_1,\dots,y_n$ are distinct elements of $\mathbb F_q$.
\begin{remark}
    The codes generated by $G_1$ and $G_3$ are MDS if and only if $\sum_{i\in I} y_i\notin \{0, 1\}$, for every subset $I\subset [n]$ of size $k-1$, whereas the code generated by $G_2$ is MDS if and only if $\sum_{i\in I} y_i\notin \{1, \lambda\}$, for every such subset. The codes generated by $G_i$, $i=1,2,3$ can be viewed as variants of Roth--Lempel codes. Similar constructions have also been considered in \cite{li2025new,sui2027new}. However, these variants do not yield MDS codes of lengths exceeding the maximum length attainable by Roth--Lempel codes.
\end{remark}    
\subsection{CTRS Codes via Points on Real Secants}
In this subsection, we consider $[n+1,k]_q$ and $[n+2,k]_q$ CTRS codes whose projective systems consist of $n$ points on $C_{k-1}$ and the remaining point(s) on a real secant of $C_{k-1}$.

For $1\le n\le q+1,$ let $\bm{x}:=(x_1,x_2, \dots, x_{n})$ be an ordered $n$-tuple of distinct elements of $\mathbb{F}_q\cup \{\infty\}$, and let $\lambda\in \mathbb{F}_q$. For $a,b\in (\mathbb{F}_q\cup \{\infty\})\setminus\{x_i : 1\le i \le n\}$ with $a\ne b$, define
    \begin{equation}
        G_k(\bm{x}, a, b, \lambda):=\begin{bmatrix}
            c_{k-1}(x_1) & c_{k-1}(x_2) & \cdots & c_{k-1}(x_n) & c_{k-1}(a)-\lambda c_{k-1}(b)
        \end{bmatrix}.
    \end{equation}
    For $\lambda_1, \lambda_2 \in \mathbb{F}_{q}$ with $\lambda_1 \lambda_2\ne 1$, define
    \begin{equation}
        G_k(\bm{x}, a, b, \lambda_1, \lambda_2):=\begin{bmatrix}
            c_{k-1}(x_1) & \cdots & c_{k-1}(x_n) & c_{k-1}(a)-\lambda_1 c_{k-1}(b) & \lambda_2 c_{k-1}(a)-c_{k-1}(b)
        \end{bmatrix}.
    \end{equation}
\begin{remark}
     The CTRS codes generated by $G_k(\bm{x},a,b,\lambda)$ and $G_k(\bm{x},a,b,\lambda_1,\lambda_2)$ were studied in \cite{Liu2026}. There is a minor difference in our definition of $G_k(\bm{x}, a, b, \lambda_1, \lambda_2)$ --  the definition in \cite{Liu2026} uses $(c_{k-1}(a)-\lambda_2 c_{k-1}(b))$ as 
     the last column of $G_k(\bm{x}, a, b, \lambda_1, \lambda_2)$, where as we use  $(\lambda_2 c_{k-1}(a)-c_{k-1}(b))$, so as to   allow  both $c_{k-1}(a)$ and $c_{k-1}(b)$ to be columns of $G_k(\bm{x}, a, b, \lambda_1, \lambda_2)$. It was shown in \cite{Liu2026} that the code generated by $G_k(\bm{x},a,b,\lambda)$ is MDS if and only if 
     \[
        \lambda\ne \prod_{\substack{i\in I\\ x_i\ne\infty}}\frac{a-x_i}{b-x_i},
     \]
     for every subset $I\subset [n]$ of size $k-1$. Similarly, it was shown that the code generated by $G_k(\bm{x},a,b,\lambda_1,\lambda_2)$ is MDS if and only if
\[
    \lambda_1\ne \prod_{\substack{i\in I\\ x_i\ne\infty}}\frac{a-x_i}{b-x_i}
    \quad\text{and}\quad
    \lambda_2\ne \prod_{\substack{i\in I\\ x_i\ne\infty}}\frac{b-x_i}{a-x_i}
\]
for every subset $I\subset[n]$ with $|I|=k-1$. Moreover, it was shown that the code generated by $G_k(\bm{x},a,b,\lambda)$ is non-GRS whenever $\lambda\ne 0$ and $3\le k\le (n+1)/2$. Similarly, the code generated by $G_k(\bm{x},a,b,\lambda_1,\lambda_2)$ is non-GRS whenever at least one of $\lambda_1$ and $\lambda_2$ is non-zero and $3\le k\le (n+2)/2$. In fact, we show that these codes are non-GRS for a broader range. Up to monomial equivalence, it suffices to consider the codes generated by $G_k(\bm{x},0,\infty,\lambda)$ and $G_k(\bm{x},0,\infty,\lambda_1,\lambda_2)$, where $\bm{x}$ is an ordered $n$-tuple of distinct elements of $\mathbb{F}_q^{\times}$.
\end{remark}
\begin{proposition}\label{prop: nonGRS of real chords}

 The code generated by $G_k(\bm{x},0,\infty,\lambda)$ (with $\lambda\ne 0$) is  non-GRS not only for $3\le k \le (n+1)/2$ but also for the broader range $3\le k\le n-2$.\\
 Similarly, the code generated by  $G_k(\bm{x},0,\infty,\lambda_1,\lambda_2)$ (with at least one $\lambda_1, \lambda_2$ non-zero)
  is  non-GRS not only for $3\le k \le (n+2)/2$ but also for the broader range $3\le k\le n-2$.
\end{proposition}
\begin{proof}
    Since $n \geq k+2$, by  Corollary \ref{corr: geometric proof idea},  the code generated by $G_k(\bm{x},0,\infty,\lambda)$ is GRS if and only if $\lambda=0$, and the  the code generated by $G_k(\bm{x},0,\infty,\lambda_1,\lambda_2)$ is GRS if and only if $\lambda_1=\lambda_2=0$. \end{proof}   
    



For odd $q$, Liu et. al in \cite{Liu2026} constructed non-GRS $[N,k]_q$-MDS codes for $N\le (q+3)/2$ and $3\le k\le N/2$. Moreover, by Proposition \ref{prop: nonGRS of real chords}, for odd $q$, one obtains non-GRS $[N,k]_q$ MDS codes for $N\le (q+3)/2$ and $3\le k\le N-4$. Liu et al. \cite{Liu2026} also provided a parity-check matrix for the code generated by $G_k(\bm{x}, a, b, \lambda_1, \lambda_2)$.
In Corollary \ref{corr: real secant parity-check two twists}, we show that the dual code of the code generated by $G_k(\bm{x}, a, b, \lambda_1, \lambda_2)$ has the structure of (row)-twisted GRS code.
\subsection{CTRS Codes via Points on Imaginary Secants}
    In this subsection, we introduce $[n+1,k]_q$ and $[n+2,k]_q$ CTRS codes whose projective systems consist of $n$ points on $C_{k-1}$ and the remaining point(s) on an imaginary secant of $C_{k-1}$. We establish necessary and sufficient conditions for such codes to be MDS. By determining the dimensions of their Schur squares, we show that these codes are non-GRS $[N, k]_q$-MDS codes for $3\le k\le N/2$. Moreover, using a geometric argument, we show that these codes are non-GRS $[N, k]$-MDS for $3\le k \le N-4$, thereby extending the range of $k$ for which non-equivalence to GRS codes can be established beyond what follows from the Schur-square argument. Finally, for odd $q$, we construct non-GRS $[N, k]_q$-MDS codes for $N\le (q+5)/2$ and $3\le k \le N-4$.

    For $1\le n\le q+1,$ let $\bm{x}:=(x_1,x_2, \dots, x_{n})$ be an ordered $n$-tuple of distinct elements of $\mathbb{F}_q\cup \{\infty\}$. Consider the quadratic extension $\mathbb{F}_{q^2}/\mathbb{F}_q$ and fix $\alpha\in \mathbb{F}_{q^2}\setminus \mathbb{F}_q$ and $\lambda\in \mathbb{F}_{q^2}^\times$. Define
    \begin{equation}
        G_k(\bm{x}, \alpha, \lambda):=\begin{bmatrix}
            c_{k-1}(x_1) & c_{k-1}(x_2) & \cdots & c_{k-1}(x_n) & \lambda c_{k-1}(\alpha)+\lambda^{q} c_{k-1}(\alpha^q)
        \end{bmatrix}.
    \end{equation}
    For $\lambda_1, \lambda_2 \in \mathbb{F}_{q^2}^{\times}$ with $\lambda_1/\lambda_2 \notin \mathbb{F}_q^\times$, we  define
    \begin{equation}
    \resizebox{\textwidth}{!}{$
        G_k(\bm{x}, \alpha, \lambda_1, \lambda_2):=\begin{bmatrix}
            c_{k-1}(x_1) & \cdots & c_{k-1}(x_n) & \lambda_1 c_{k-1}(\alpha)+\lambda_1^{q} c_{k-1}(\alpha^q) & \lambda_2 c_{k-1}(\alpha)+\lambda_2^{q} c_{k-1}(\alpha^q)
        \end{bmatrix}.
    $}
    \end{equation}
We note that the matrices $G_k(\bm{x}, \alpha, \lambda)$ and $G_k(\bm{x}, \alpha, \lambda_1, \lambda_2)$ have entries in $\mathbb{F}_q$. The next lemma provides necessary and sufficient conditions for the codes generated by $G_k(\bm{x}, \alpha, \lambda)$ and $G_k(\bm{x}, \alpha, \lambda_1, \lambda_2)$ to be MDS.
\begin{lemma}\label{lemma: NASC for CTRS imaginary chords to be MDS}
\begin{enumerate}
\item[(a)] The code $\mathcal{C}_k(\bm{x},\alpha,\lambda)$ generated by
$G_k(\bm{x},\alpha,\lambda)$ is a $[n+1,k]_q$-MDS code if and only if, for every $I\subset[n]$ with $|I|=k-1$,
\[
\Tr\left(
\lambda\!\!
\prod_{\substack{i\in I\\x_i\ne\infty}}
(\alpha-x_i)
\right)\ne0.
\]
\item[(b)] The code $\mathcal{C}_k(\bm{x},\alpha,\lambda_1,\lambda_2)$ generated by
$G_k(\bm{x},\alpha,\lambda_1,\lambda_2)$ is a $[n+2,k]_q$-MDS code if and only if, for every $I\subset[n]$ with $|I|=k-1$,
            \[
                \Tr\left(
                \lambda_j\!\!\prod_{\substack{i\in I\\x_i\ne\infty}}(\alpha-x_i)
                \right)\ne0, \qquad\text{for } j=1,2.
            \]
\end{enumerate}
\end{lemma}
\begin{proof}
   We prove only part $(b)$. Part $(a)$ follows by considering the corresponding subcases of the argument given below. Recall that a $k$-dimensional linear code is MDS if and only if every set of $k$ columns of its generator matrix is linearly independent. Thus, to prove $b)$, it suffices to show that any $k\times k$ submatrix of $G_k(\bm{x}, \alpha, \lambda_1, \lambda_2)$ is nonsingular.
   
    We divide the $k\times k$ submatrices into three cases according to the number of the last two columns that they contain.

    If the submatrix contains neither of the last two columns, then it invertible as it is a Vandermonde matrix.
    
    
    We next suppose that the submatrix contains exactly one of the last two columns. Let $I=\{i_1, i_2, \dots, i_{k-1}\}\subset [n]$ of size $k-1$. Then the corresponding submatrix is
    $$
        D_1=\begin{bmatrix}
        c_{k-1}(x_{i_1}) & \cdots & c_{k-1}(x_{i_{k-1}}) &  \lambda c_{k-1}(\alpha)+ \lambda^q c_{k-1}(\alpha^q)
    \end{bmatrix},
    $$
    where $\lambda \in \{\lambda_1, \lambda_2\}$. If $\infty \notin \{x_i  \colon i \in I\}$, then 
    \begin{align*}
        \det D_1&=\lambda\Delta_k(x_{i_1}, \dots, x_{i_{k-1}}, \alpha)+\lambda^q\Delta_k(x_{i_1}, \dots, x_{i_{k-1}}, \alpha^q)\\
        &=\Delta_{k-1}(x_{i_1}, \dots, x_{i_{k-1}})\left[\lambda \prod_{i\in I} (\alpha-x_i)+\left(\lambda \prod_{i\in I} (\alpha-x_i)\right)^q\right]\\
        &=\Delta_{k-1}(x_{i_1}, \dots, x_{i_{k-1}})\Tr\left( \lambda \prod_{i\in I} (\alpha-x_i)\right),
    \end{align*}
    which is non-zero if and only if $\Tr\left( \lambda \prod_{i\in I} (\alpha-x_i)\right)\ne 0$. 

    If $\infty \in \{x_i  \colon i \in I\}$, we may assume without loss of generality that $x_{i_{k-1}}=\infty$. In this case 
    $$
        \det D_1= \Delta_{k-2}(x_{i_1}, \dots, x_{i_{k-2}})\Tr\left(\lambda_1\prod_{i\in I\setminus\{i_{k-1}\}}(\alpha-x_i)\right),
    $$
    which is non-zero precisely under the trace condition in $b)$.
    

    Lastly, suppose that the submatrix contains both of the last two columns. Let $I=\{i_1, i_2,$  $\dots, i_{k-2}\}\subset [n]$ of size $k-2$.  The corresponding submatrix is
    $$
        D_2=\begin{bmatrix}
        c_{k-1}(x_{i_1}) & \cdots & c_{k-2}(x_{i_{k-2}}) &  \lambda_1 c_{k-1}(\alpha)+ \lambda_1^q c_{k-1}(\alpha^q) & \lambda_2 c_{k-1}(\alpha)+ \lambda_2^q c_{k-1}(\alpha^q)
    \end{bmatrix}.
    $$
    By the linearity of the determinant in the last two columns, 
    \begin{align*}
    \det D_2&=\lambda_1 \lambda_2^q \Delta_{k}(x_{i_1}, \dots, x_{i_{k-2}}, \alpha, \alpha^q)+\lambda_1^q \lambda_2 \Delta_{k}(x_{i_1}, \dots, x_{i_{k-2}}, \alpha^q, \alpha)\\
    &= (\lambda_1 \lambda_2^q -\lambda_1^q \lambda_2) \Delta_{k}(x_{i_1}, \dots, x_{i_{k-2}}, \alpha, \alpha^q).
    \end{align*}

      The quantity $\Delta_{k}(x_{i_1}, \dots, x_{i_{k-2}}, \alpha, \alpha^q)$ is non-zero as 
    $x_{i_1}, \dots, x_{i_{k-2}}, \alpha, \alpha^q$ are distinct. Since ${\lambda_1}/{\lambda_2}\notin \mathbb{F}_q^\times$, the expression $(\lambda_1 \lambda_2^q -\lambda_1^q \lambda_2)$ is non-zero.  Consequently, $\det D_2\ne 0$.
    This completes the proof.
\end{proof}
We now present explicit constructions of the codes $\mathcal{C}_k(\bm{x}, \alpha, \lambda)$ and $\mathcal{C}_k(\bm{x}, \alpha, \lambda_1, \lambda_2)$ that are MDS and are inequivalent to any GRS code.
\begin{theorem}\label{thm: construct using mult subgroup}
    Let $H$ be a subgroup of $\ker(\Nr)$, and let $\alpha\in\mathbb{F}_{q^2}\setminus\mathbb{F}_q$. Let $h_1, \dots, h_n$  be distinct elements of $H$, and define
    \[
        x_i=
            \begin{cases}
            \dfrac{\alpha^q h_i-\alpha}{h_i-1},& h_i\ne1,\\[2mm]
            \infty,& h_i=1.
            \end{cases}
    \]
    \begin{enumerate}
        \item [(a)] If $-\lambda^{q-1}\notin H,$ then the code $\mathcal{C}_k(\bm{x}, \alpha, \lambda)$ generated by $G_k(\bm{x}, \alpha, \lambda)$ is a non-GRS $[n+1, k]_q$-MDS code for $3\le k \le n-2$.
        \item [(b)] If $-\lambda_j^{q-1}\notin H,$ for $j=1,2,$ then the code $\mathcal{C}_k(\bm{x}, \alpha, \lambda_1, \lambda_2)$ generated by $G_k(\bm{x}, \alpha, \lambda_1, \lambda_2)$ is a non-GRS $[n+2, k]_q$-MDS code for $3\le k \le n-2$. 
        
    \end{enumerate}
  
\end{theorem}
\begin{proof}

The element 
\[
    g(t)=\frac{\alpha^q t-\alpha}{t-1}
\]
of $\pgl_2(\mathbb{F}_{q^2})$ maps  $\ker(\Nr)$ bijectively to  $\mathbb{F}_q\cup\{\infty\}$. Since the elements $h_i$ are pairwise distinct and belong to $\ker(\Nr)$,
the evaluation points $x_i=g(h_i)$ are pairwise distinct elements of $\mathbb{F}_q\cup\{\infty\}$.

We now use Lemma~\ref{lemma: NASC for CTRS imaginary chords to be MDS}(a) to show that the code $\mathcal{C}_k(\bm{x},\alpha,\lambda)$ is MDS. Suppose, to the contrary, that there exists $I\subset[n]$ with $|I|=k-1$ such that
$$
    \Tr\left(\lambda\!\!\prod_{\substack{i\in I\\x_i\ne\infty}}(\alpha-x_i)\right)=0.
$$
Then
$$
    \lambda\prod_{\substack{i\in I\\x_i\ne\infty}}
    (\alpha-x_i)+\lambda^q\prod_{\substack{i\in I\\x_i\ne\infty}}(\alpha^q-x_i)=0,
$$
or, equivalently,
\begin{equation}\label{eqn: lambda condition for one CTRS via imaginary chord}
-\lambda^{q-1}=\prod_{\substack{i\in I\\x_i\ne\infty}}
\left(\frac{\alpha-x_i}{\alpha^q-x_i}\right).
\end{equation}
For each $i\in I$ with $x_i\ne\infty$, we have $x_i=\dfrac{\alpha^qh_i-\alpha}{h_i-1},$ and hence $\dfrac{\alpha-x_i}{\alpha^q-x_i}=h_i.$

Substituting this in \eqref{eqn: lambda condition for one CTRS via imaginary chord}, we obtain
$$
-\lambda^{q-1}=\prod_{\substack{i\in I\\x_i\ne\infty}}h_i \in H,
$$
which contradicts the hypothesis.
Hence, by Lemma \ref{lemma: NASC for CTRS imaginary chords to be MDS}(a), the code $\mathcal{C}_k(\bm{x},\alpha,\lambda)$ is a $[n+1,k]_q$-MDS code.

We next determine the dimension of the Schur square of  $\mathcal{C}_k(\bm{x},\alpha,\lambda)$. For $1\le i\le k,$ let
$$
    \bm{g}_i:=\left(x_1^{i-1}, x_2^{i-1}, \dots, x_n^{i-1}, \lambda \alpha^{i-1}+\lambda^q \alpha^{(i-1)q}\right)
$$
denote the $i$th row of $G_k(\bm{x},\alpha,\lambda)$. Then $\mathcal{C}_k(\bm{x},\alpha,\lambda)^{\star 2}$ is generated by 
$$
    \mathcal{D}:=\{\bm{g}_i\star \bm{g}_j: \ 1\le i \le j\le k\}.
$$
For $1\le j\le k,$ we have $\bm{g}_1\star \bm{g}_j=\left(x_1^{j-1}, x_2^{j-1}, \dots, x_n^{j-1},\ * \right)$, while, for $2\le i \le k,\ $$\bm{g}_i\star \bm{g}_k=\left(x_1^{i+k-2}, x_2^{i+k-2}, \dots, x_n^{i+k-2},\ * \right).$

In addition, $\bm{g}_2\star \bm{g}_2=\left(x_1^2, x_2^2, \cdots, x_n^2, (\lambda \alpha+ (\lambda \alpha)^q)^2 \right).$ Its first $n$ coordinates coincide with those of $\bm{g}_1\star\bm{g}_3$, whereas their last coordinates differ. Thus,
$$
    \bm{g}_2\star \bm{g}_2- \bm{g}_1\star \bm{g}_3=\left(0, 0, \cdots, 0, -\lambda^{q+1}(\alpha-\alpha^q)^2 \right).
$$
Since $\lambda\ne0$ and $\alpha\notin\mathbb{F}_q$, we have $-\lambda^{q+1}(\alpha-\alpha^q)^2\ne0.$ Moreover, for $1\le i\le j\le k$ and $1\le s\le t\le k$ satisfying
$i+j=s+t$, we have
$$
    \bm{g}_i\star \bm{g}_j-\bm{g}_s\star\bm{g}_t=\left(0, \cdots, 0, -\lambda^{q+1}\left(\alpha^{s-1}\alpha^{q(t-1)}+\alpha^{t-1}\alpha^{q(s-1)}-\alpha^{i-1}\alpha^{q(j-1)}-\alpha^{j-1}\alpha^{q(i-1)}\right) \right).
$$
Hence,
$$
\begin{aligned}
&\bm{g}_i\star\bm{g}_j-\bm{g}_s\star\bm{g}_t\\
&\quad-
\frac{
\alpha^{s-1}\alpha^{q(t-1)}+\alpha^{t-1}\alpha^{q(s-1)}-\alpha^{i-1}\alpha^{q(j-1)}-\alpha^{j-1}\alpha^{q(i-1)}
}{(\alpha-\alpha^q)^2}
\left(
\bm{g}_2\star\bm{g}_2-\bm{g}_1\star\bm{g}_3
\right)
=\bm{0}.
\end{aligned}
$$
Thus, $\bm{g}_i\star\bm{g}_j-\bm{g}_s\star\bm{g}_t$ is a scalar multiple of $\bm{g}_2\star\bm{g}_2-\bm{g}_1\star\bm{g}_3$ whenever $i+j=s+t.$

Consequently, after elementary row operations, a generator matrix for $\mathcal{C}_k(\bm{x},\alpha,\lambda)^{\star 2}$ can be written in the form
$$
\begin{bmatrix}
    c_{2k-2}(x_1) & c_{2k-2}(x_2) & \cdots & c_{2k-2}(x_n) & *\\
    0 & 0 & \cdots & 0 & -\lambda^{q+1}(\alpha-\alpha^q)^2
\end{bmatrix}.
$$
The rank of the above matrix is $\min\{2k, n+1\}$, since the last row is linearly independent of the first $2k-1$ rows. Consequently, $\dim\left(
\mathcal{C}_k(\bm{x},\alpha,\lambda)^{\star 2}
\right)=\min\{2k, n+1\}.$ Using Remark \ref{remark: non-RS condition}, for $3\le k\le (n+1)/2$, the code $\mathcal{C}_k(\bm{x},\alpha,\lambda)^{\star 2}$ is inequivalent to any GRS code.

We now use Corollary \ref{corr: geometric proof idea} to show that the code $\mathcal{C}_k(\bm{x},\alpha,\lambda)$ is not equivalent to any GRS code for $3\le k\le n-2$. It suffices to show that  $\bm{u}=\lambda c_{k-1}(\alpha)+\lambda^{q} c_{k-1}(\alpha^q)$ does not represent a point of $C_{k-1}$. Indeed, if $\bm{u} = a c_{k-1}(\beta)$ for some $a \in \mathbb F_q$, then $c_{k-1}(\alpha), c_{k-1}(\alpha^q), c_{k-1}(\beta)$ would be linearly dependent over $\mathbb{F}_{q^2}$, which is not the case.




The proof of part $(b)$ follows analogously.
\end{proof}
\begin{corollary}\label{corr: codes of length q+5/2}
\begin{enumerate}
    \item [(a)] Let $q$ be an odd prime power. Then, for every $k\ge 3$ and  $k+4\le N\le \frac{q+5}{2}$, there exists a non-GRS $[N,k]_q$-MDS code.
    \item [(b)] Let $q=2^{2m+1}$, $m\ge 1$. Then, for every $k\ge 3$ and  $k+4\le N\le \frac{q+10}{3}$, there exists a non-GRS $[N,k]_q$-MDS code.
\end{enumerate}
\end{corollary}
\begin{proof}
 (a). Let $\gamma$ be a primitive element of $\mathbb{F}_{q^2}$. Then $\ker(\Nr)=\langle \gamma^{q-1} \rangle$.  Let $H:=\langle \gamma^{2(q-1)}\rangle$. Then $H$ is a subgroup of $\ker(\Nr)$ of order $(q+1)/2$.
    We claim that $\lambda=\gamma^{\frac{q+3}{2}}$ satisfies $-\lambda^{q-1}\notin H$. Suppose, to the contrary, that $-\lambda^{q-1}\in H$. Since $-1=\gamma^{\frac{q^2-1}{2}}$ and every element of $H$ is of the form $\gamma^{2(q-1)j}$ for some $0\le j<\frac{q+1}{2}$, we have
\[
    \gamma^{\frac{q^2-1}{2}}\gamma^{\frac{(q-1)(q+3)}{2}}= \gamma^{2(q-1)j}
\]
for some $0\le j<\frac{q+1}{2}$. This yields $\gamma^{(q-1)(q+2)}=\gamma^{2(q-1)j}$. Since $\gamma$ has order $q^2-1=(q-1)(q+1)$, it follows that $q+2\equiv 2j\pmod{q+1}$, a contradiction, since $q$ is odd. Hence, $-\lambda^{q-1}\notin H.$

Selecting $\lambda=\gamma^{\frac{q+3}{2}}$ and choosing the evaluation points $x_1, \dots, x_{N-1}$, where $N-1\le |H|$, according to Theorem \ref{thm: construct using mult subgroup}, all the conditions of Theorem \ref{thm: construct using mult subgroup}(a) are satisfied. Hence, we obtain a non-GRS $[N,k]_q$-MDS code with
    \[
        N\le |H|+1=\frac{q+3}{2}\qquad\text{and}\qquad 3\le k \le N-3.
    \]

Let $\lambda_1=\gamma^{\frac{q+3}{2}}$ and $\lambda_2=\gamma^{\frac{q+7}{2}}$. Proceeding as above, we get $-\lambda_1^{q-1},-\lambda_2^{q-1}\notin H.$ Moreover, $\frac{\lambda_1}{\lambda_2}=\gamma^{-2}\notin\mathbb{F}_q^{\times}.$ Indeed, if $\gamma^{-2}\in\mathbb{F}_q^\times$, then $\gamma^{-2(q-1)}=1,$ which would imply $(q+1)\mid2$, a contradiction since $q$ is odd. Choosing $\lambda_1=\gamma^{\frac{q+3}{2}}$, $\lambda_2=\gamma^{\frac{q+7}{2}}$ and the evaluation points $x_1, \dots, x_{N-2}$, where $N-2\le |H|$, according to Theorem \ref{thm: construct using mult subgroup}, all the conditions of Theorem \ref{thm: construct using mult subgroup}(b) are satisfied. Hence, we obtain a non-GRS $[N, k]_q$-MDS code with 
    \[
        N\le |H|+2=\frac{q+5}{2}\qquad\text{and}\qquad 3\le k \le N-4.
    \]
    (b). Let $\gamma$ be a primitive element of $\mathbb{F}_{q^2}$. Set $K:=\ker(\Nr)$ and set $\beta:=\gamma^{q-1}$. Then $K=\langle \beta \rangle$.  Let $H:=\langle \beta^3 \rangle.$ Since $q=2^{2m+1},$ $3\mid q+1$. Consequently, $H$ is a subgroup of $K$ of order $\frac{q+1}{3}$ and $[K: H]=3.$ Hence, $K=H \sqcup \beta H \sqcup \beta^2 H$. Consider the evaluation set $\bm{x}=(x_1, \dots, x_{|H|}, x_{|H|+1}),$ where $x_i$ are defined as in Theorem \ref{thm: construct using mult subgroup}, for $1\le i\le |H|$ and $x_{|H|+1}=\frac{\alpha^q \beta-\alpha}{\beta-1}.$ Following the proof of Theorem \ref{thm: construct using mult subgroup}, the code $\mathcal{C}_k(\bm{x}, \alpha, \lambda_1, \lambda_2)$ is MDS if $\lambda_j^{q-1}\notin H\cup \beta H,$ for $j=1,2.$

    Let $\lambda_1=\gamma^2$ and $\lambda_2=\gamma^5$. Then $\lambda_1^{q-1}=\beta^2\in \beta^2 H$ and $\lambda_2^{q-1}=\beta^5\in \beta^2 H$. Thus, $\lambda_j^{q-1}\notin H\cup \beta H,$ for $j=1,2$. Moreover, $\frac{\lambda_1}{\lambda_2}=\gamma^{-3}\notin \mathbb{F}_q^{\times}$. Indeed if $\gamma^{-3}\in \mathbb{F}_q^{\times}$, then $(q+1)\mid 3$, which is a contradiction, since $q=2^{2m+1}$, $m\ge 1$. Choosing $\lambda_1=\gamma^2$, $\lambda_2=\gamma^5$, and the evaluation points $x_1, \dots, x_{N-2}$, where $N-2\le |H|+1$, yields a non-GRS $[N, k]_q$-MDS code with 
    \[
        N\le (|H|+1)+2=\frac{q+10}{3}\qquad\text{and}\qquad 3\le k \le N-4.
    \]
\end{proof}
\begin{remark}
For odd $q$, the construction of Liu et al. in \cite{Liu2026} produces non-GRS MDS codes over $\mathbb{F}_q$ of length at most $\frac{q+3}{2}$, whereas our construction yields non-GRS MDS codes over $\mathbb{F}_q$ of length at most $\frac{q+5}{2}$. For $q=2^{2m}$, Liu et al. in \cite{Liu2026} obtain non-GRS MDS codes over $\mathbb{F}_q$ of length at most $\frac{q+8}{3}$, while for $q=2^{2m+1}$, our construction yields non-GRS codes over $\mathbb{F}_q$ of length at most $\frac{q+10}{3}$.
\end{remark}
\section{Dual of CTRS Codes}\label{section 4}
In this section, we determine the dual of an arbitrary CTRS code and show that its dual is a (row-)twisted GRS code.

Consider the elementary symmetric polynomial $e_r(x_1, x_2, \dots, x_n)$. If $x_j=\infty$ for some $1\le j \le n,$ we define $e_r(x_1, x_2, \dots, x_n):=e_r(x_1, \dots, \hat{x_j}, \dots, x_n),$ where $\hat{x_j}$ indicates that $x_j$ is omitted. Throughout this section, we write
\[
    e_r:=e_r(x_1, x_2, \dots, x_n)\quad\text{and}\quad h_r:=h_r(x_1, x_2, \dots, x_n).
\]
    Define the lower triangular matrix
    \begin{equation}
        L_k:=\left[(-1)^{i-j} e_{i-j}\right]_{1\le i, j\le k}.
    \end{equation}
    Consider the matrix
    \[
        G_1=\begin{bmatrix}
                    c_{k-1}(x_1) & c_{k-1}(x_2) & \cdots & c_{k-1}(x_n) & \bm{u}
                \end{bmatrix},
    \]
    where $\bm{u}=(u_0, u_1, \dots, u_{k-1})\in\mathbb{F}_q^k$.
    The following lemma gives an explicit description of the dual of the code generated by $G_1$. Lemma \ref{lemma: parity-check one twist} also follows from \cite[Equation (8)]{kaipa2017deep}. For completeness, we provide an independent proof.
\begin{lemma}\cite{kaipa2017deep}\label{lemma: parity-check one twist}
    Let
    \begin{equation*}
        f_{\bm{u}}(X):=X^{m-k}(a_1 X^{k-1}+a_2 X^{k-2}+\cdots+ a_{k-1}X+a_k),
    \end{equation*}
    where
    \[
        m:=|\{x_1, x_2, \dots, x_n\}\setminus \{\infty\}|\quad\text{and}\quad[a_1\ a_2\ \cdots\ a_k]^\top=L_k \bm{u}.
    \]
    Set $f_{\bm{u}}(\infty):=0$. Then the dual of the code generated by $G_1$ is the (row-)twisted GRS code generated by
    \begin{equation}
        H_1=\begin{bmatrix}
            c_{n-k-1}(x_1) & c_{n-k-1}(x_2) & \cdots & c_{n-k-1}(x_n) & \bm{0}\\
            f_{\bm{u}}(x_1) & f_{\bm{u}}(x_2) & \cdots & f_{\bm{u}}(x_n) & 1
            \end{bmatrix}\textnormal{diag}\left(\delta_1, \cdots, \delta_n, -1\right),
    \end{equation}
    where $\bm{0}$ is the $(n-k)\times 1$ zero vector, and for $1\le i\le n,$
    \[
        \delta_i=
        \begin{cases}
        \displaystyle\prod_{\substack{j=1\\ j\ne i,\ x_j\ne\infty}}^n
        \!\!\!\!\!\!(x_i-x_j)^{-1}, & \text{if } x_i\ne\infty,\\[3mm]
        -1, & \text{if } x_i=\infty.        \end{cases}
    \]
\end{lemma}
\begin{proof}
 We first consider the case when each $x_i\in \mathbb{F}_q$. There exist $v_1,\dots,v_n\in\mathbb{F}_q$ such that
\begin{equation}\label{eqn: existence of v_i}
    \sum_{j=1}^n \delta_j v_j c_{k-1}(x_j)=\bm{u},
\end{equation}
since the matrix $[\delta_1c_{k-1}(x_1)\ \cdots \ \delta_n c_{k-1}(x_n)]$ has rank $k$.
Then
\[
H=
\begin{bmatrix}
c_{n-k-1}(x_1)&c_{n-k-1}(x_2)&\cdots&c_{n-k-1}(x_n)&\bm{0}\\
v_1&v_2&\cdots&v_n&1
\end{bmatrix}
\operatorname{diag}(\delta_1,\dots,\delta_n,-1)
\]
is a parity-check matrix of the code generated by $G_1$. Indeed,
$G_1H^\top=O$, and the rows of $H$ are linearly independent.

        Define the Lagrange interpolation polynomial $u(X)$ satisfying $u(x_i)=v_i$ for all $i\in [n]$ by
        \[
            u(X):=\sum_{i=1}^nv_i
            \prod_{\substack{j=1\\j\ne i}}^n
            \left(\frac{X-x_j}{x_i-x_j}\right)=\sum_{i=1}^n\delta_i v_i
            \prod_{\substack{j=1\\j\ne i}}^n(X-x_j).
        \]
        We first claim that
        \begin{equation}\label{eqn: matrix form of u(X)}
            u(X)=
            [X^{n-1}\ \cdots\ X \ 1]
            L_n
            [\delta_1 c_{n-1}(x_1)\ \cdots \ \delta_n c_{n-1}(x_n)]
            [v_1 \ \cdots \ v_n]^\top.
        \end{equation}
        Indeed, the right-hand side can be written as 
        \[
        \sum_{i=1}^n\delta_i v_i
            [X^{n-1}\ \cdots\ X \ 1]
            L_n c_{n-1}(x_i).
        \]
        By the definition of $L_n$, the $r$-th entry of $L_n c_{n-1}(x_i)$ is
        \[
        \sum_{s=1}^{r}(-1)^{r-s}e_{r-s}x_i^{s-1}
        =\sum_{t=0}^{r-1}(-1)^t e_t x_i^{r-1-t},
        \]
        for $1\le r\le n$. Consequently,
    \[
        [X^{n-1}\ \cdots\ X \ 1]
        L_n c_{n-1}(x_i)
        =\sum_{r=1}^{n}\left(\sum_{t=0}^{r-1}(-1)^t e_t x_i^{r-1-t}\right)X^{n-r}
        =\prod_{\substack{j=1\\j\ne i}}^n(X-x_j).
    \]
This proves the claim. Next note that, due to \eqref{eqn: existence of v_i},
\[
    [X^{n-1}\ \cdots\ X^{n-k}]L_k [\delta_1 c_{k-1}(x_1)\ \cdots \ \delta_n c_{k-1}(x_n)][v_1 \ \cdots \ v_n]^\top=[X^{n-1}\ \cdots\ X^{n-k}]L_k \bm{u}.
\]
Therefore, the coefficients of $X^{n-1}, \dots, X^{n-k}$ in $u(X)$ are given by
\[
[a_1\ a_2\ \cdots\ a_k]^\top=L_k\bm u,
\]
Consequently,
\[
u(X)=
f_{\bm u}(X)+r(X),
\qquad \deg r\le n-k-1.
\]
Thus,
\[
    H=\begin{bmatrix}
            c_{n-k-1}(x_1) & c_{n-k-1}(x_2) & \cdots & c_{n-k-1}(x_n) & \bm{0}\\
           f_{\bm u}(x_1)+r(x_1) & f_{\bm u}(x_2)+r(x_2) & \cdots & f_{\bm u}(x_n)+r(x_n) & 1
            \end{bmatrix}\textnormal{diag}\left(\delta_1, \cdots, \delta_n, -1\right),
\]
Since $\deg r\le n-k-1$, the polynomial $r(X)$ can be expressed as a linear combination of $1,X,\ldots,X^{n-k-1}$. Hence, by subtracting a suitable linear combination of the first $(n-k)$ rows of $H$, we obtain the parity-check matrix 
\begin{equation}
        H_1=\begin{bmatrix}
            c_{n-k-1}(x_1) & c_{n-k-1}(x_2) & \cdots & c_{n-k-1}(x_n) & \bm{0}\\
            f_{\bm{u}}(x_1) & f_{\bm{u}}(x_2) & \cdots & f_{\bm{u}}(x_n) & 1
            \end{bmatrix}\textnormal{diag}\left(\delta_1, \cdots, \delta_n, -1\right).
    \end{equation}
Next, assume without loss of generality that $x_n=\infty$. By interchanging the last two columns of $G_1$, we get $\widetilde{G_1}=[c_{k-1}(x_1) \ \cdots \ c_{k-1}(x_{n-1}) \ \bm{u} \ c_{k-1}(\infty)]$. Let $G_1':=[c_{k-1}(x_1) \ \cdots \ c_{k-1}(x_{n-1}) \ \bm{u}]$. By the preceding argument with $n$ replaced by $n-1$, a parity-check matrix of the code generated by $G_1'$ is
\[
    H_1'=\begin{bmatrix}
            c_{n-k-2}(x_1)  & \cdots & c_{n-k-2}(x_{n-1}) & \bm{0}\\
            f_{\bm{u}}(x_1)  & \cdots & f_{\bm{u}}(x_{n-1}) & 1
            \end{bmatrix}\textnormal{diag}\left(\delta_1, \cdots, \delta_{n-1}, -1\right).
\]
Consequently, a parity-check matrix of the code generated by $\widetilde{G_1}$ is 
\[
    \begin{bmatrix}
            c_{n-k-2}(x_1) &  \cdots & c_{n-k-2}(x_{n-1}) & \bm{0} & \bm{0}\\
            f_{\bm{u}}(x_1)  & \cdots & f_{\bm{u}}(x_{n-1}) & 1 & 0\\
            x_1^{n-k-1}  & \cdots & x_{n-1}^{n-k-1} & 0 & 1
            \end{bmatrix}\textnormal{diag}\left(\delta_1, \cdots, \delta_{n-1}, -1, -1\right).
\]
Interchanging the last two rows yields another parity-check matrix:
\[
    \widetilde{H_1}=\begin{bmatrix}
            c_{n-k-1}(x_1) &\cdots & c_{n-k-1}(x_{n-1}) & \bm{0} & c_{n-k-1}(\infty)\\
            f_{\bm{u}}(x_1)  & \cdots & f_{\bm{u}}(x_{n-1}) & 1 & 0
            \end{bmatrix}\textnormal{diag}\left(\delta_1, \cdots, \delta_{n-1}, -1, -1\right).
\]
Finally, interchanging the last two columns of $\widetilde{H_1}$ gives a parity-check matrix of the code generated by $G_1$:
\[
   H_1=\begin{bmatrix}
            c_{n-k-1}(x_1) &\cdots & c_{n-k-1}(x_{n-1})  & c_{n-k-1}(\infty) & \bm{0}\\
            f_{\bm{u}}(x_1)  & \cdots & f_{\bm{u}}(x_{n-1}) & f_{\bm{u}}(\infty) & 1
            \end{bmatrix}\textnormal{diag}\left(\delta_1, \cdots, \delta_{n-1}, \delta_n, -1\right),
\]
where
\[
    f_{\bm{u}}(X):=X^{n-k-1}(a_1 X^{k-1}+a_2 X^{k-2}+\cdots+ a_{k-1}X+a_k)\quad\text{and}\quad f_{\bm{u}}(\infty)=0.
\]
This completes the proof.
\end{proof}
Consider the matrix 
    $$
        G_2=\begin{bmatrix}
                c_{k-1}(x_1) & \cdots & c_{k-1}(x_n) & \bm{u} &\bm{v}
            \end{bmatrix},
    $$
where $\bm{u}=(u_0, u_1, \dots, u_{k-1})$ and $\bm{v}=(v_0, v_1, \dots, v_{k-1})\in \mathbb{F}_q^k$. The following lemma gives an explicit description of the dual of the code generated by $G_2.$
\begin{lemma}\label{lemma: parity-check double twist}
    With the notations as in Lemma \ref{lemma: parity-check one twist}, the dual of the code generated by $G_2$ is the (row-)twisted GRS code generated by
    $$
        H_2=\begin{bmatrix}
            c_{n-k-1}(x_1) & c_{n-k-1}(x_2) & \cdots & c_{n-k-1}(x_n) & \bm{0} & \bm{0}\\
            f_{\bm{u}}(x_1) & f_{\bm{u}}(x_2) & \cdots & f_{\bm{u}}(x_n) & 1 & 0\\
            f_{\bm{v}}(x_1) & f_{\bm{v}}(x_2) & \cdots & f_{\bm{v}}(x_n) & 0 & 1\\
            \end{bmatrix}\textnormal{diag}\left(\delta_1, \cdots, \delta_n, -1, -1\right).
    $$
    \end{lemma}
    \begin{proof}
        The proof follows directly from the proof of Lemma \ref{lemma: parity-check one twist}, with the same argument applied to both $\bm{u}$ and $\bm{v}$.
    \end{proof}
\begin{corollary}
    For $k\le n\le q,$ let $\bm{x}:=(x_1, x_2, \dots, x_n)$ be an ordered $n$-tuple of distinct elements of $\mathbb{F}_q$ and $\lambda\in \mathbb{F}_q$. Then the dual of the Roth--Lempel code generated by
     $$
        \begin{bmatrix}
            c_{k-1}(x_1) & \cdots & c_{k-1}(x_n) & c_{k-1}(\infty) &\lambda c_{k-1}(\infty)+c_{k-1}'(\infty)
        \end{bmatrix}
    $$
    is the (row-)twisted GRS code generated by
    $$
        \begin{bmatrix}
            c_{n-k-1}(x_1) & \cdots & c_{n-k-1}(x_n) & c_{n-k-1}(\infty) &\bm{0}\\
            f(x_1) & \cdots & f(x_n) & 0 & 1
        \end{bmatrix}\textnormal{diag}\left(\delta_1, \cdots, \delta_n, -1\right),
    $$
    where
    $$
        f(X)=\left(\lambda-\sum_{i=1}^n x_i\right)X^{n-k}+X^{n-k+1},\quad \text{and}\quad \delta_i=\prod_{\substack{j=1\\ j\ne i}}^n
        (x_i-x_j)^{-1}.
    $$

\end{corollary}
\begin{proof}
    Using the notations as in Lemma \ref{lemma: parity-check one twist}, we have $\bm{u}=(0, 0, \dots, 0, 1, \lambda)$. Therefore, 
    $$
       a_i=i\text{th\ entry of\ }L_k \bm{u}=\sum_{j=1}^{i} (-1)^{i-j} e_{i-j} u_{j-1}= \begin{cases}
           0, & \text{for\ }i\le k-2\\
           1, & \text{for\ }i=k-1\\
           -e_1+\lambda, & \text{for\ }i=k
       \end{cases}.
    $$
    Hence, 
    $$
        f(X):=f_{\bm{u}}(X)=\sum_{i=1}^{k} a_i X^{n-i}=\left(\lambda-\sum_{i=1}^n x_i\right)X^{n-k}+X^{n-k+1}.
    $$
\end{proof}
\begin{corollary}\label{corr: real secant parity-check one twist}
    For $k\le n\le q-1,$ let $\bm{x}:=(x_1, x_2, \dots, x_n)$ be an ordered $n$-tuple of distinct elements of $\mathbb{F}_q^{\times}$ and $\lambda\in \mathbb{F}_q^{\times}$. Then the dual of the code generated by 
    $$
        G_k(\bm{x}, 0, \infty, \lambda):=\begin{bmatrix}
            c_{k-1}(x_1) & c_{k-1}(x_2) & \cdots & c_{k-1}(x_n) & c_{k-1}(0)-\lambda c_{k-1}(\infty)
        \end{bmatrix}
    $$
    is the (row-)twisted GRS code generated by
    $$
        \begin{bmatrix}
            c_{n-k-1}(x_1) &  c_{n-k-1}(x_2) & \cdots & c_{n-k-1}(x_n) & \bm{0}\\
            f(x_1) & f(x_2) & \cdots & f(x_n) & 1
        \end{bmatrix}\textnormal{diag}\left(\delta_1, \cdots, \delta_n, -1\right),
    $$
    where $$
    f(X)=\sum_{i=0}^{k-1} (-1)^{i} e_{i} {X}^{n-i-1}-\lambda  X^{n-k},\quad \text{and}\quad \delta_i=\prod_{\substack{j=1\\ j\ne i}}^n
        (x_i-x_j)^{-1}.
    $$
\end{corollary}
\begin{proof}
    Using the notations as in Lemma \ref{lemma: parity-check one twist}, we have $\bm{u}=(1, 0, \dots, 0, -\lambda)$. Therefore, 
    $$
       a_i=i\text{th\ entry of\ }L_k \bm{u}=\sum_{j=1}^{i} (-1)^{i-j} e_{i-j} u_{j-1}= \begin{cases}
           (-1)^{i-1}e_{i-1}, & \text{for\ }i<k\\
           (-1)^{i-1}e_{i-1}-\lambda, & \text{for\ }i=k
       \end{cases}.
    $$
    Hence, 
    $$
        f(X):=f_{\bm{u}}(X)=\sum_{i=1}^{k} a_i X^{n-i}=\sum_{i=0}^{k-1} (-1)^{i} e_{i} {X}^{n-i-1}-\lambda {X}^{n-k}.
    $$
\end{proof}
\begin{corollary}\label{corr: real secant parity-check two twists}
    For $k\le n\le q-1,$ let $\bm{x}:=(x_1, x_2, \dots, x_n)$ be an ordered $n$-tuple of distinct elements of $\mathbb{F}_q^{\times}$ and $\lambda_1\ne \lambda_2\in \mathbb{F}_q^{\times}$. Then the dual of the code generated by 
    \begin{equation*}
    \resizebox{\textwidth}{!}{$
        G_k(\bm{x}, 0, \infty, \lambda_1, \lambda_2):=\begin{bmatrix}
            c_{k-1}(x_1) & \cdots & c_{k-1}(x_n) & c_{k-1}(0)-\lambda_1 c_{k-1}(\infty) & \lambda_2 c_{k-1}(0)-c_{k-1}(\infty)
        \end{bmatrix}
        $}
    \end{equation*}
    is the (row-)twisted GRS code generated by
    $$
        \begin{bmatrix}
            c_{n-k-1}(x_1) &  \cdots & c_{n-k-1}(x_n) & \bm{0} & \bm{0}\\
            f_1(x_1) & \cdots & f_1(x_n) & 1 & 0\\
            f_2(x_1) & \cdots & f_2(x_n) & 0 & 1
        \end{bmatrix}\textnormal{diag}\left(\delta_1, \cdots, \delta_n, -1, -1\right),
    $$
    where $$
    f_1(X)=\sum_{i=0}^{k-1} (-1)^{i} e_{i} {X}^{n-i-1}-\lambda_1 {X}^{n-k},
    $$
    $$
        f_2(X)=\lambda_2\sum_{i=0}^{k-1} (-1)^{i} e_{i} {X}^{n-i-1}- {X}^{n-k}\quad\text{and}\quad\delta_i=\prod_{\substack{j=1\\ j\ne i}}^n
        (x_i-x_j)^{-1}.
    $$
\end{corollary}
\begin{proof}
    Follows immediately by applying Lemma \ref{lemma: parity-check double twist} and adapting the argument in the proof of Corollary \ref{corr: real secant parity-check one twist}.
\end{proof}
For a polynomial $f(X)=\sum_{i=0}^{k-1} a_i X^{i}\in \mathbb{F}_{q^2}[X]$, define $f^{[q]}(X):=\sum_{i=0}^{k-1} a_i^q X^{i}$.
\begin{corollary}\label{corr: img secant parity-check one twist}
    For $k\le n\le q+1,$ let $\bm{x}:=(x_1,x_2, \dots, x_{n})$ be an ordered $n$-tuple of distinct elements of $\mathbb{F}_q\cup \{\infty\}$ and let $\alpha\in \mathbb{F}_{q^2}\setminus \mathbb{F}_q$ and $\lambda\in \mathbb{F}_{q^2}^\times$. Then the dual of the code generated by
    $$
        G_k(\bm{x}, \alpha, \lambda):=\begin{bmatrix}
            c_{k-1}(x_1) & c_{k-1}(x_2) & \cdots & c_{k-1}(x_n) & \lambda c_{k-1}(\alpha)+\lambda^{q} c_{k-1}(\alpha^q)
            \end{bmatrix}
    $$
    is the (row-)twisted GRS code generated by
    $$
        \begin{bmatrix}
            c_{n-k-1}(x_1) &  c_{n-k-1}(x_2) & \cdots & c_{n-k-1}(x_n) & \bm{0}\\
            f(x_1) & f(x_2) & \cdots & f(x_n) & 1
        \end{bmatrix}\textnormal{diag}\left(\delta_1, \cdots, \delta_n, -1\right),
    $$
    where $$
    f(X)=g(X)+g^{[q]}(X),\quad g(X)=\lambda\sum_{i=1}^{k}\left(\sum_{j=1}^{i} (-1)^{i-j} e_{i-j} \alpha^{j-1}\right) {X}^{n-i}\quad\text{and} 
    $$
    $\delta_i$ is defined as in Lemma \ref{lemma: parity-check one twist}.
\end{corollary}
\begin{proof}
    Using the notations as in Lemma \ref{lemma: parity-check one twist}, we have $\bm{u}=(\lambda+\lambda^q, \lambda\alpha+(\lambda\alpha)^q, \dots, \lambda\alpha^{k-1}+(\lambda\alpha^{k-1})^q)$. Therefore, 
    $$
       a_i=i\text{th\ entry of\ }L_k \bm{u}=\sum_{j=1}^{i} (-1)^{i-j} e_{i-j} u_{j-1}=\sum_{j=1}^{i}(-1)^{i-j} e_{i-j} (\lambda\alpha^{j-1}+(\lambda\alpha^{j-1})^q).
    $$
    Hence, 
    \begin{align*}
        f(X):=f_{\bm{u}}(X)=\sum_{i=1}^{k} a_i X^{n-i}&=\sum_{i=1}^{k} \left(\sum_{j=1}^{i}(-1)^{i-j} e_{i-j} (\lambda\alpha^{j-1}+(\lambda\alpha^{j-1})^q)\right) {X}^{n-i}\\
        &=g(X)+g^{[q]}(X),
    \end{align*}
    where $g(X)=\lambda\sum_{i=1}^{k}\left(\sum_{j=1}^{i} (-1)^{i-j} e_{i-j} \alpha^{j-1}\right) {X}^{n-i}.$
\end{proof}
\begin{corollary}
    For $k\le n\le q+1,$ let $\bm{x}:=(x_1,x_2, \dots, x_{n})$ be an ordered $n$-tuple of distinct elements of $\mathbb{F}_q\cup \{\infty\}$ and let $\alpha\in \mathbb{F}_{q^2}\setminus \mathbb{F}_q$ and $\lambda_1, \lambda_2\in \mathbb{F}_{q^2}^\times$. Then the dual of the code generated by
    \begin{equation*}
    \resizebox{\textwidth}{!}{$
        G_k(\bm{x}, \alpha, \lambda_1, \lambda_2):=\begin{bmatrix}
            c_{k-1}(x_1) & \cdots & c_{k-1}(x_n) & \lambda_1 c_{k-1}(\alpha)+\lambda_1^{q} c_{k-1}(\alpha^q) & \lambda_2 c_{k-1}(\alpha)+\lambda_2^{q} c_{k-1}(\alpha^q)
        \end{bmatrix}.
    $}
    \end{equation*}
    is the (row-)twisted GRS code generated by
    $$
        \begin{bmatrix}
            c_{n-k-1}(x_1) &  \cdots & c_{n-k-1}(x_n) & \bm{0} & \bm{0}\\
            f_1(x_1) & \cdots & f_1(x_n) & 1 & 0\\
            f_2(x_1) & \cdots & f_2(x_n) & 0 & 1
        \end{bmatrix}\textnormal{diag}\left(\delta_1, \cdots, \delta_n, -1, -1\right),
    $$
    where $$
    f_1(X)=\lambda_1 g(X)+\lambda_1^q g^{[q]}(X),\quad f_2(X)=\lambda_2 g(X)+\lambda_2^q g^{[q]}(X),
    $$
    $$ g(X)=\sum_{i=1}^{k}\left(\sum_{j=1}^{i} (-1)^{i-j} e_{i-j} \alpha^{j-1}\right) {X}^{n-i}
    $$
    and $\delta_i$ is defined as in Lemma \ref{lemma: parity-check one twist}.
\end{corollary}
\begin{proof}
    Follows immediately by applying Lemma \ref{lemma: parity-check double twist} and adapting the argument in the proof of Corollary \ref{corr: img secant parity-check one twist}.
\end{proof}
\section{Conclusion}\label{Section 5}
    In this article, we considered codes whose projective systems consist of points on a rational normal curve together with at most two points on an imaginary secant of the curve. We derived necessary and sufficient conditions for these codes to be MDS and provided explicit constructions. Using a geometric approach, we proved that every $[n,k]_q$ MDS CTRS code is non-GRS for $3\le k\le n-4$, thereby extending the range of dimensions for which non-GRS property can be established beyond the limitations of the Schur-square technique. In particular, for odd $q$, we constructed non-GRS MDS codes of length $(q+5)/2$ for every dimension $k$ satisfying $3\le k\le (q-3)/2$. Finally, we determined the structure of the dual of an arbitrary CTRS code and showed that its dual is a (row-)twisted GRS code.
\section*{Declarations}
\subsection*{Conflict of Interest}
Both authors declare that they have no conflict of interest.

\subsection*{Acknowledgments}
The first author acknowledges the support of the Anusandhan National Research Foundation (ANRF) under the National Post-Doctoral Fellowship scheme, File No. PDF/2026/001259.

\bibliographystyle{amsplain}
\bibliography{CTRS}

\end{document}